# A Re-evaluation of Shor's Algorithm

John W. Cooper*

IPST, University of Maryland, College Park MD, 20754

ABSTRACT

Shor's algorithm, which outlines a method of factoring large numbers using quantum computation has played a vital part in establishing quantum computation as an active area of research in physics and computer science. It is one of the few quantum algorithms whose end result is obtained in numerical form. While a number of authors have written explaining the algorithm and some authors have doubted it's usefulness there has been few attempts to examine the algorithm critically from the standpoint of it's eventual use to obtain the factors of large numbers. That is the purpose of the present report.

* E-mail address: jwc@umd.edu

# I. Introduction

The publication in 1994 by Peter Shor of a quantum computational algorithm for factoring integers (1) was met by both positive and negative reactions. Berthiaume and Brassard (2) added the following note in proof to their paper on Oracle quantum computing. "Our conclusion that 'a good way of pumping funding into the building of an actual quantum computer would be to find an efficient quantum factoring algorithm!' was prophetic. Peter Shor discovered such an algorithm within hours of our writing this sentence and sending the final version of our paper to the Editors." Svizol on the other hand presented a short paper (3) which argued that although an exponential speedup was in principal correct for quantum computation any such speedup would be accompanied by an attenuation of detection rates.

Shor's work has had possibly a greater effect on the development of the field of quantum computing than any other factor. Several authors have written detailed explanations of the algorithm either in complete articles (4-8) or as an important part of books (9-13) or review articles (14-17) devoted to quantum computing. Research has been devoted to effectively improving the algorithm using alternative strategies (18-20) and to studying the role of entanglement (21) in the use of the algorithm. Classical computing models (22-23) have been developed to apply the method to the problem of factoring small integers and experimentally the smallest integer that can be factored via the method,15, was successfully factored via an NMR experiment (24). Finally, practically every paper written on quantum computing acknowledges Shor's work as an example of the value of quantum computation.

# II. Number Theoretical Basis of the Algorithm

The basic computation used in the algorithm is the evaluation of the function:
$$f(x) = A^x \bmod N \qquad (1)$$
where N is the positive integer to be factored, A is a positive integer less than N which is not a factor of N and contains no common factors with N and x is a positive integer variable which runs from 0 to M where $N^2 < M < 2N^2$.

If we were to calculate

$$f(x) = A^x \qquad (2)$$
directly, the function f(x) would be "one to one"; i.e., for every value of x there would be a unique value of f(x). This would also be true if x were a continuous positive real variable rather than an integer. In general any function evaluated mod N will not be "one to one". If, for example,
$$f(x) = f(x+c) \qquad (3)$$
where c is a real positive constant, the function will be "two to one". If, on the other hand.

$$f(x)= f(x\pm c) \qquad (4)$$

the function is no longer "two to one" since $x\pm c$ can assume negative values for $c>x$ and in general $f(-x)\neq f(x)$. For example, for the function of Eq.2 $f(-x)= 1/f(x)$. This is because there are now 2 values of c in Eq. 3; $+c$ and $-c$.

Even if x and c are limited to positive real variables, any function of a single variable can in general be r to 1. In this case:

$$F(x)= f(x+c_i) \qquad (5)$$

where in general there can be r different values of $c_i$.

We now give two simple examples of periodic functions which are pertinent to the analysis of the Shor algorithm. The first is the function

$$g(\delta)=De^{i\delta} =D(\cos\delta+i\sin\delta) \qquad (6)$$

where D and $\delta$ are real and positive.

The important point here is that $g(\delta)=De^{i\delta}$ mod $2\pi$ so that it may be written in the form of Eq. 4, i.e.;

$$g(\delta)= -g(\delta\pm\pi) \qquad (7)$$

i.e., the function is mod $2\pi$ but both real and imaginary parts change sign when $\pi$ is added or subtracted. Looking at Eq. 3, 4 and 6 we see that as we have expanded the definition of our variable from positive quantities to both positive and negative quantities and finally to both real and imaginary positive and negative quantities the periodicity changes. $g(\delta)$ may be real or imaginary, positive or negative or any combination of these.

The second is any positive integer expressed in binary notation; i.e.;

$$B= a_0 2^0+a_1 2^1+ a_2 2^2+ a_3 2^3+ a_4 2^4+ a_5 2^5+ \cdots \qquad (8)$$

Here the $a_j$'s form a string of 0's and 1's. Then the function B mod 2 is "two to one"; i.e., it can only have the values 1 or 0. Similarly B mod 3 is "three to one" and in general B mod M is "M to one".

The number of different values that can occur for a function that is finite over a given interval is called the order of the function; i.e., the order of a two to one function is 2, a three to one function has order 3 and in general an rth to one function has order r.

In Eq.1 $f(x)$, x, N and A are all positive integers. Under these circumstances, the order r of $f(x)$ (an integer) is given simply by the equation:

$$A^r=1 \bmod N \qquad (9)$$

From Eq. 8, if the maximum integer that can be expressed is M (corresponding to $M=2^L-1$ and all a"s equal to 1) then the order of B is M: i.e., we can count up to M and then we start over. For Eq. 9 the order r depends on A, N but does not depend on M, the maximum allowed value of x in Eq. 1. Alternatively r is the number of different values of x for which $A^x$ in Eq 1 is less than N. Clearly, r<N and will decrease as A increases.

One might wonder why the detailed discussion of "the meaning of mod" given above is included since all of the information given is common knowledge to most undergraduate students of mathematics and (hopefully) also to most computer scientists. The reason is that it provides a clear definition of the information content of the fundamental building blocks of classical and quantum computing where numbers are involved. Each term in Eq 8 corresponds to a binary bit and is completely specified by its location j and the coefficient $a_i$ which can be 0 or 1. Eq. 6, on the other hand specifies a single qubit in terms of a positive amplitude D and a positive phase $\delta$.* As with classical bits each term in a string of qubits can be specified by a location label j.

If r is even and N is an odd integer, then it is possible to find factors of N from a knowledge of r. This is based on the fact that for r even

$$(A^{r/2}-1)(A^{r/2}+1) = 0 \text{ Mod } N \tag{10}$$

This implies that $A^{r/2} = \pm 1$. Since only one of factors on the left can be zero, implying that Eq. 10 is valid, the other factor divided by N must be an integer.

The procedure for finding r consists of evaluating Eq. 1 using a quantum computer for all integer values of x<N and then inferring r from the results of the calculation. The procedure will not work in all cases. In particular it will fail if r turns out to be odd or if it turns out that $A^r = -1$ Mod N, since then $A^{r/2}$ Mod N is imaginary.

The analysis of the algorithm in terms of number theory was originally given by Shor (1) and has been repeated and elaborated on by a number of authors; e.g.,( 4-12 ). The main object of this analysis is to estimate the probability of success in obtaining factors via the computation.

*This of course is not the most general definition of a qubit or the one used by most workers in the field. However, it has the advantage of simplicity and is adequate for the present discussion.

# III. The Basic Steps in the Algorithm

## A. Defining the Machine

The first step in evaluating the usefulness of the algorithm is to define the type of computer that is to be used to do the computation. This is important since in principle the algorithm could be run on any of the estimated 938 million digital computers in the world today successfully. The problem is that although this is true it could not achieve useful results for factoring large numbers in the foreseeable future. The same is true if a number of the computers acted in parallel at the same time to do parts of the computation. A quantum computer can in principle run the algorithm much faster and produce reasonable results.

The type of computer necessary has been described by Shor (1) and more detailed requirements for a general purpose computer capable of doing numerical computations has been given by DeVincenzo ( 25). The necessary requirements are:

A. A number of quantum registers, each of which consists of qubits of the form of Eq. 8.Each register is ordered in a similar fashion to Eq. 8.Mathematically the form of the register is obtained by replacing each of the "0" 's of Eq.8 by $\cos\delta_j$ , the "1" 's by $i\sin\delta_j$ and D by $1/\sqrt{L}$ where L is the number of qubits in the register.
( Note that this definition of a register is somewhat different than those usually given since, as noted above, each qubit depends only on a single phase shift. However it is adequate for the purposes of defining the steps needed to realize the algorithm, since with this definition each qubit can be any vector in the complex plane.The normalization is important since it insures that the sum of the number of states in the register must add up to unity.)

B. One must be able to initiate the machine; i.e., for each register to set each of the $\delta_j$ of the L qubits in a given register to predefined values.

C. A quantum network capable of performing the calculation and storing the result in one of the quantum registers must be available. The process must be reversible; i.e. , if one started the computer in its final state , one must be able to run the calculation backwards and reach the initial state.

D. A method of extracting the answer from the machine.

A great deal of thought has been given to the necessary steps that must be taken to successfully design a machine satisfying criteria A and C and quantum networks have been designed ( 26,27 ) which will successfully do all of the steps necessary in the implementation of Shor's algorithm. Criteria B and D have not been studied as intensively and, as we shall see pose difficulties to the practical application of the algorithm.

## B.  Initiation

In order to perform the calculation  4 registers are needed and we define them as follows:

I(x):            Input register which contains the values of x
F(f(x)):         Output register which contains the computed values of f(x)
A  and  N:     Input registers which contains the numerical values of A and N

The input register is first set to zero. What this means is that the phases of each of the qubits in the register  must be set to zero. The output register is also set to zero. The A and N registers are set to the numerical values of  A and N  in Eq. 1; i.e.,  if A is represented as a binary  string , the phase shifts of the elements of the string corresponding to  "0" are set to 0 and those corresponding to  "1" are set to $\pi/2$.( This description  of  initiation differs somewhat from many descriptions given in the literature. Many discussions ,including  Shor's,  consider only two registers  and set them both to zero.)

Mathematically  the above procedure seems simple enough  but we must consider what it means in terms of a physical system. A qubit is envisioned  as a physical system which can only have two states (0> and 1>) but at any given time is not in either one of them. The amplitudes corresponding to these states  are $\pm\cos\delta$ and $\pm i\sin\delta$. In general we don't know the phase but we do know that the physical system is in one state or the other.The probability of it's being  in 0>  is then $\cos^2\delta$  and of it's being in 1>  is $\sin^2\delta$. Note that these definitions can be switched ; i.e., 0> corresponds to $\pm\cos\delta$ and  1> to $\pm i\sin\delta$ .

It is worthwhile noting that the same is true  for classical bits such as those which comprise the string represented by Eq. 8.If  we reverse every bit in the string we of course obtain the compliment which if a given string represents A modN can be interpreted as –A mod N. Thus it can be said that classical computing works in two orthogonal spaces "0"  (or "-1") and "1". Similarly quantum computing works in four orthogonal spaces "1"."-1" , "i" and "-i".

A word needs to be said about registers A and N.  In  Shor's  original  work  it is not mentioned as it is assumed that they are built into the system. However, since A is to be changed if the algorithm fails to produce a result  and we may wish to factor some other  integer there must be some means of  doing that rather than rebuilding the quantum computer. The registers containing A and N  need not be quantum registers  since they are never changed during a particular computation so they could be envisioned as read only memory . However, they have to interact with the  quantum mechanical registers.

Mathematically, the problem of initiation is simple. One simply sets all of the phases of the individual qubits to zero. For a physical system this is not so straightforward. The standard way to do this is to create a given physical system by some procedure (say an atom in its ground state), perform a measurement on it to ascertain what state it was in before the measurement and then  recreate it again by  the same procedure. This has been done for a single qubit in an ion trap and has been done for several qubits in NMR experiments  via the technique of  creating effectively pure states. How this is to be achieved for the larger number of qubits needed to factor large numbers may be a problem.

# C. Calculation of Eq. 1

The calculation of Eq. 1 via a quantum mechanical network consisting of a number of qubits and quantum mechanical gates which cause the computer to evolve from its initial state following initiation to a final state from which the periodicity of Eq. 1 can be extracted has been described numerous times in the literature. Detailed networks have been presented by a number of groups and the methods used have been discussed in detail in the literature. Essentially they follow the procedures outlined in Shor's original paper. Here we will only give an outline of the method and refer the reader to Refs. 26 and 27 for more detailed accounts.

First consider how one would compute the function given by Eq.1 classically. One would load a register with a value of x, compute $f(x) = A^x$ Mod N and store it in a second register. The sequence of operations would be:

$$(0)(0) \rightarrow (x)(0) \rightarrow (x)f(x)$$

A quantum computation works in the same manner but with two important exceptions. First, as mentioned previously, the process must be reversible, meaning that in principle one could run the calculation backwards, taking $(x)f(x)$ as the initial state of the computer and obtain $(0)(0)$ as the final state. Second, and most important, it is possible to put the initial state of the first register into a state which is the linear combination of 0 and of all of the integers between 0 and $M=2^L$ where L is the number of qubits in the register. The initial state would then be $\frac{1}{\sqrt{M}} \sum_{j=0}^{M-1} j$. This can be done simply by rotating each qubit through an angle of π/4.

The first step in the algorithm after initiation is simply this. The input register is put into a state which is a linear combination of 0 and all of the integers between 0 and M.

The second step is the most detailed one in the algorithm and requires a quantum network of gates. Specifically it computes the function of Eq 1, stores it in the output register and "cleans up" any part of the computation that if left intact would mean that the calculation could not be run in the reverse direction.

In order to describe this step and the following steps we introduce the following notation following Ref. 13. The initial state of the computer is written as:

$$|\Psi_1\rangle = 2^{-(m+n)/2} |0\rangle_0 |0\rangle_1 \cdots |0\rangle_{n-1} |0\rangle_0 |0\rangle_1 \cdots |0\rangle_{m-1} \qquad (11)$$

indicating that the both the input register containing n qubits and the output register containing m qubits have been set to zero. The labeling of the qubits indicates the power of two associated with the qubit.

When each qubit in the input register is rotated by π/4 the input register has the form:

$$I = \frac{1}{\sqrt{M}} (|0\rangle + |1\rangle)_0 (|0\rangle + |1\rangle_1) \cdots (|0\rangle + |1\rangle)_{n-1} \qquad (12)$$

where M =$2^n$. This can be written as:

$$I = \frac{1}{\sqrt{M}} \sum_{j=0}^{M-1} |j\rangle \qquad (13)$$

and the state of the computer as:

$$\Psi_2 = \frac{1}{\sqrt{M}} \sum_{j=0}^{M-1} |j\rangle|0\rangle \qquad (14)$$

Next Eq. 1 is evaluated and the result stored in the output register. The state of the computer is now:

$$\Psi_3 = \frac{1}{\sqrt{M}} \sum_{j=0}^{M-1} |j\rangle|A^j \bmod N\rangle \qquad (15)$$

This can be written as:

$$\Psi_3 = \frac{1}{\sqrt{M}} \sum_{l=0}^{r-1} \sum_{q=0}^{s_l} |qr+l\rangle|A^l \bmod N\rangle \qquad (16)$$

where r is the periodicity we seek and $s_l$ is the largest integer for which $s_l r+l<M$. Note that there will be r different values of l and consequently r values of $A^l$ modN.

Finally a quantum Fourier transform is applied to the input register. The state of the machine is now:

$$\Psi_4 = \frac{1}{\sqrt{M}} \sum_{l=0}^{r-1} \sum_{q=0}^{s_l} \frac{1}{\sqrt{n}} \sum_{p=0}^{M-1} e^{\frac{2\pi i p(qr+l)}{n}} |p\rangle|A^l \bmod N\rangle \qquad (17)$$

When this step is completed the calculation is done and all that remains to do is to extract relevant information from the machine to determine if possible r, the order of $A^l$ mod N, and if it is possible, to obtain factors of N from this information and, if not, repeat a similar calculation in order to obtain more information. There are various schools of thought as to how this is to be done which we will discuss in detail. First, however we consider what is the input and desired output of the calculation. The input is simple. It consists of real positive integers A and N. However, there is another implied input to the calculation, namely n and m, the sizes of the input and output registers.

This is important since as has been stressed many times the computational complexity of the problem depends on the number of qubits needed. Vedral et al.(25) have given a simple estimate of the number of qubits needed in general to perform the algorithm. If s qubits are needed to encode the number to be factored N, then the total number needed for the algorithm is estimated as 7s+1. This number can be reduced depending upon how the computational networks are set up but will be increased if additional qubits are needed for error correction or to increase the relative accuracy of the computed results.

## IV. Measurement; Analysis of Simple Cases

In order to see how the results of the computation are obtained via measurement it is useful to consider the much discussed simplest example of factoring 3x5 = 15. Following Refs.11 and 12 we represent the input and output registers by 4 qubits each; i.e, n=m =4. The state of the input register just before evaluating Eq. 1 will be given by Eq. 13, the superposition of 0 and the first 15 integers with an overall normalization factor of ¼. Evaluating Eq. 1 using A=7 results in four different values of F(x), namely 1,7,4 and 13. Writing this state of the machine in decimal notation we have:

$$\Psi_3 = \frac{1}{4}((|0\rangle+|4\rangle+|8\rangle+|12\rangle)|1\rangle) \qquad (18)$$

$$+(|1\rangle+|5\rangle+|9\rangle+|13\rangle)|7\rangle)$$

$$+(|2\rangle+|6\rangle+|10\rangle+|14\rangle)|4\rangle)$$

$$+(|3\rangle+|7\rangle+|11\rangle+|15\rangle)|13\rangle))$$

Performing the Fourier transform we have:

$$\Psi_4 = \frac{1}{4}((|0\rangle+|4\rangle+|8\rangle+|12\rangle)|1\rangle) \qquad (19)$$

$$+(|0\rangle+i|4\rangle-|8\rangle-i|12\rangle)|7\rangle)$$

$$+(|0\rangle-|4\rangle+|8\rangle-|12\rangle)|4\rangle)$$

$$+(|0\rangle-i|4\rangle-|8\rangle+i|12\rangle)|13\rangle))$$

One reason for writing this out is to call attention to the fact that performing the Fourier transform actually enhances the probability of measuring a particular value in the input register. If we measure after the transform we can only obtain the values 0,4, 8 and 12, each with probabilities ¼. If we measured before the transform we could obtain any value from 0 to 15 with probability 1/16 for each value. Also note that each of the values 0,4,8 and 12 is associated with (entangled) all of the values of the output register. Thus it appears that, contrary to many of the explanations of the algorithm, a measurement of the output register actually reduces the probability of successfully obtaining one of the possible values of the input register. In the above example , measurement of the output register would yield either 1,7,4 or 13 each with a probability of ¼.

A further measurement of the input register would then yield 0 , 4, 8 or 12 each with probability of ¼ so that the probability of getting a specific result from both measurements would be 1/16 and nothing would have been gained by performing the Fourier transform.

Now let's look at the measurement process in some detail. We have used 4 qubits for both registers. Consequently measurement of the input register is expected to yield the following results each with probability of ¼ .

$$
\begin{array}{ll}
0000 & (\ 0) \\
0100 & (\ 4) \\
1000 & (\ 8) \\
1100 & (12)
\end{array}
\qquad (20)
$$

The pattern shown in Eq. 20 is general for any periodicity which has the form $r= 2^K$ . As shown in reference 26 , if this is the case and the number of qubits in the input register is L  ($n= 2^L$), then the result of Fourier transforming is a bit pattern where the first L- K bits are zero and the next L bits are as likely to be "0" or "1".Had we used 5 qubits rather than 4 for the input register, our results would have been 0,8,16, and 24 each with probability ¼. For this simple case we could have used only 3 qubits instead of 4 and in that case the final result would have been either 0,2, 4 and 6 each with probability of ¼.

The above example is instructive since it allows us to view the calculation in a different way. We can consider an N fold group of integers as a set of Dirac delta functions $\delta(x)$ , x=1,N lying along the x axis.The result of our calculation can then be view as simply a rescaling of the discrete variable x by multiplying each x by r where r is the periodicity. Viewed in another way this rescaling means that the distance between the Dirac delta functions which was unity before the calculation is now r so that the transformation can be written as $\delta(x) \to \delta(xr)$. For each delta function its distance from the origin will be nr where n is an integer. One can simply take the analytic Fourier transform of each delta function which will be simply $\delta(1/nr)$.

A more complicated example of this is the factorization of 119 (=7x17) which has recently been studied in Ref. 21 to determine the role of entanglement in the algorithm. Using A=92 the period is found to be 16 (K=4).Using 13 qubits for the input register , the 16 values of the input register that could be measured would lie at 0, 1024, 2048…..15,360 each with a probability of 1/16. The corresponding bit patterns expected to be observed would be:

0000000000000    (1024x0)

```
0001000000000   (1024x1)
0010000000000   (1024x2)
0011000000000    etc.                              (21)
0100000000000
0101000000000
0110000000000
0111000000000
1000000000000
1001000000000
1010000000000
1011000000000
1100000000000
1101000000000
1110000000000
1111000000000
```

Now let's see how we would obtain the periodicity from a single run of the algorithm which produced one of the bit patterns of Eq. 21. We would measure each qubit. If there are no errors we must obtain 0 for the first nine qubits on the right and either 0 or 1 for each of the 4 leftmost qubits. Since there are only 2 possibilities for each qubit the probability to obtain any of the 16 possible numbers is 1/16.

If our measurement produced 0 we cannot determine r. For any other value r is determined by $r=2^{L+1}\lambda/c$ where c is the measured value and $\lambda$ is an integer ranging from 1 to 15. Alternatively we could merely count the number of zeros and obtain k=13-9 and r=16.

This seems simple enough, but there is a problem of extending the procedure to factoring large numbers where r cannot be expressed as a power of 2. The problem is that for the any periodicity of the form $r=2^K$ one will obtain r different possible results and the probability of obtaining any one of them will be proportional to 1/r. Moreover, if the periodicity is not of this form, in general there will be more than r possible results and thus the probability of any given result will be less than or equal to 1/r. If r is large as it is expected to be for large N and factors of comparable size it is unlikely that one could obtain useful results from a single run of the algorithm.

Of course, as was pointed out in Ref. 21 for the case where $r=2^K$ one doesn't need a quantum algorithm to determine the periodicity since it that case the number to be factored must be of the form $N=2^M+1$ and r can be found simply by dividing by powers of 2.

In the analysis of the algorithm it is often said that the probability of obtaining a result is high. This assumption is based on the following reasoning.

Suppose, that as above, r is in fact equal to $2^K$. Then it is argued that since there are r possible answers that can be obtained via measurement it is certain that we will obtain one of them. Applying this reasoning to the above example one would obtain one of 16 possible answers and since one obviously cannot obtain the period from a measured value of zero the probability of success is 15/16.

This reasoning is incorrect as can be seen by the following example.Suppose we wish to measure the probability of elastic scattering of an electron from an atom We have good reason to assume that the scattering is isotropic; i.e., is independent of angle of ejection relative to the incident direction of the electron beam.To the extent that this is so we can determine the probability simply by forming a beam of electrons of known energy, scattering them and detecting the number of electrons we observe in a given scattering solid angle (making sure that they have not lost energy). The probability of elastic scattering will then be simply the ratio of detected to incident electrons times $4\pi$ times the solid angle. For low energies elastic scattering is expected to be the dominant scattering process occurring. If this is so the total probability of elastic scattering will be close to unity. However, if we attempt to measure it by observing within a given solid angle, the results obtained will be proportional to the solid angle used for the experiment.

The type of experiment described above which has been performed by a number of experimental physicists over the last century is completely analogous to the problem of determining a specific number using Shor's algorithm. Even though the total probability of obtaining some answer is close to unity, the probability of obtaining a specific answer is not.

There is another interesting way of looking at the result shown in Eq. 21.The results which show the final state of the input register after the algorithm has run can be compared with the initial value of that register. Looked at in this way we see that the first line corresponds to the algorithm not changing the input register at all. Also note that one can think of the input register as consisting of two parts; the right hand side which is unchanged by the algorithm and the left hand side in which each qubit has a probability of ½ of either changing or not changing. We could determine r by counting the number of qubits that can change.

Actually, the single experiment that realizes Shor's algorithm (24) does not rely on a single evaluation of the algorithm. The experiment using the NMR technique actually performs the experiment on a collection of independent molecules each of which can be viewed as a single quantum computer. Consequently, the results obtained are the same as the bitwise average of a number of runs made on a single quantum computer. The results they obtain for factoring 15 using A=7 are consistent with the analysis given above. Since only 3 qubits are used L=3 and since only one bit is zero K=2 and consequently r=4..Similarly, their results for A=11 indicate that only one qubit can change and thus r=2

*( A man with one thermometer knows the temperature. A man with two isn't sure.)*

The above quotation was one of the favorite remarks made when discussing the question of measurement by a physicist in the temperature measurement laboratory of the Bureau of Standards ( now NIST ). It is important here since it emphasizes that fact that when you measure something, your measurement is meaningless unless you also state its uncertainty. Experimentalists typically express the results of their measurements and indicate in some form the estimated uncertainty. Theorists who calculate specific results rarely do. Most of the earlier work on quantum computing tacitly assumed that measurements could be made with absolute certainty, all that one was looking for was a label for a particular quantum mechanical state. More recent work which is directed towards actually building devices that could actually carry out the calculation of a quantum mechanical algorithm provide a more detailed analysis of the measurement process.

Basically there are three ways in which errors can invalidate the results that are obtained from an evaluation of Shor's algorithm using a quantum computer, namely:
- (a) Errors in the initiation procedure which starts the computer in a well defined initial state.
- (b) Errors which occur in the evaluation of Eq. 1 or in performing the Fourier transform of Eq. 17 due to inaccuracies in the phases or amplitudes using in the various gates used to perform the computations.
- (c) Errors in the readout of the final answer of the computation.

It has been realized for some time that errors are unavoidable in any computation and much work has been devoted to developing error correcting procedures which would make it possible for computations to be successful in spite of these errors. The emphasis has so far been on developing techniques to correct errors in step (b) above, but recently the work has been extended also to steps (a) and (c). Here we will only be concerned with the effect of errors on the readout of a result..

Suppose we have in fact developed a program to evaluate Shor's algorithm and also have a quantum computer that will run the program. It is assumed that after running the program the answer is contained only in the input register defined above and we wish to measure it. In order to do this we must measure each qubit;i.e., determine whether it corresponds to 0> or 1> . As pointed out in Ref.27 , if F is the fidelity of the measurement of a single qubit; i.e., the probability of obtaining a result corresponding to either 0> or 1> the fidelity for the measurement of a register consisting of n qubits will be $F^n$. As also is pointed out in Ref. 27, the current fidelities for single qubit readout (F<.99) are not high enough to obtain meaningful results for registers containing $10^3$ qubits so that alternative procedures are being developed to increase the fidelity of single qubit readout (27,28). However, there is another problem. As noted above, the probability of obtaining a given number c as the result of a single run of the algorithm is $\leq 1/r$. What this means is that for large r we are unlikely to obtain an answer even if the readout can be accomplished with high fidelity since as r→∞ the probability of obtaining a result →0. One might think that one could improve the situation by rerunning the algorithm. While this is useful in principle to obtain more than one value of c, it does not increase the probability of obtaining any particular value. Each time the program is run some particular value of c would be obtained with a probability of $\leq 1/r$.

Since there are in general more than r values of c possible, the probability of obtaining one of them by rerunning the algorithm would be $\leq 1/r^2$.

## V. An Alternative Strategy

Clearly, the straightforward approach outlined above of performing the calculation and then making measurements on each qubit in both the input and output registers does not seem to be the best way to proceed in order to obtain factors for large numbers. As discussed above, the measurement of the output register in addition to being not necessary is actually counterproductive and the simple rerunning of the algorithm after a measurement of the input register is unlikely to yield meaningful results for large values of N. However, there are alternative procedures that may prove useful.

First, it is interesting to note that one need not perform the second quantum Fourier transform after the calculation of Eq.1 at all. The same effect can be obtained by simply setting the output register to 1. What this would do is to effectively set the value of $l = 0$ in Eq.16. Consequently if a measurement were made after doing this, the only values of x that could be obtained would be integers times r. This is exactly the same result that would be obtained by carrying out the Fourier transform and then measuring the out put register before measuring the input register. As was pointed out above one loses information by performing a measurement of the output resister but if that is to be done the alternative procedure of merely setting the output register to 1 appears to be far simpler than performing the Fourier transform and then making a measurement..

The above may be useful, but it does not solve the problem of the low probability of obtaining a specific result via a measurement of the input register after a single run of the algorithm. The following procedure could be used to alleviate this difficulty.

1. Run the algorithm as outlined above.
2. Instead of making any measurements add the contents of the input register to an auxiliary register which initially has been set to 0.
3. Rerun the algorithm.
4. In this step and all further steps add the contents of the input register to the auxiliary register.
5. Repeat steps 2-4 as many times as it is anticipated are necessary to obtain a meaningful result.
6. Measure the auxiliary register.

Once these steps have been carried out, the auxiliary register will contain an integer which will be the sum of a particular set of those integers which would have been obtained by successive runs of the algorithm. Consequently, the same procedures that were suggested for the original algorithm can be used to obtain r from the measured value of the auxiliary register.

## VI. Suggestions for Future Work

First, some work is needed to provide information on how the method would work for factoring large numbers. Since, as we have shown, the probability of success decreases as N increases there must be a limit on the size of N that can attempted to be factored using any given machine and quantum program. This subject has not been addressed.

Second and equally important some consideration should be given to the lower limit of the usefulness of the algorithm. It is quite obvious that using the algorithm to factor 15 is inefficient compared to doing the computation classically. If in fact the method is more efficient for large numbers it would be useful to know how large a number must be in order for the method to have an advantage.